# Positron Channeling

K.S. Badikyan

National University of Architecture and Construction, Yerevan, Armenia

badikyan.kar@gmail.com

## Abstract

The possibility of channeling the low-energy relativistic positrons around separate crystallographic axes with coaxial symmetry of negative ions in some types of crystals is shown. The process of annihilation of positrons with electrons of medium was studied in detail.

## 1. Introduction

The study of the methods for generation of shortwave coherent radiation was always one of essential problems of science stimulated by its application possibilities. The mechanism of radiation of relativistic electrons at their motion in periodical structures was first proposed in [1] (see also [2]), the structure of this type being termed the undulator. Owing to this publication [1] the process of creation of modern generation devices, the free electron lasers (FEL) [3, 4], was substantially promoted. Despite the fact that the generation technology of undulator radiation (UR) is successfully developed and its achievements are evident [5–7], some problems nevertheless still exist and require solution up to the present. The UR frequency is determined by the length of periodical structure (for FEL with an undulator it is a macroscopic device, ranging from several millimeters to several centimeters), the energy of electrons and the magnetic field.
The disadvantages of undulators are the large size (a few tens of meters) and use of high energy electrons (on the order of GeV and higher).

After discovery of electron (positron) channeling in crystals [8–10] and of the accompanying short-wave radiation a prospect appeared that above problems may be solved even partially. Particularly, the problems of short-wave X-ray radiation from lower energy electrons (on the order of several MeV) at very short distances (of about several micrometer) have been solved. However, there arose a new problem connected with the fact that in the channeling regime the lifetime of particles (electrons and positrons) was usually vary small , that was too short for transformation of the substantial part of particle energy to the radiation energy.

Many authors were involved in the development of the quantum theory of electrons and positrons channeling [11–13]. It should be observed that an electron in crystal may undergo both the planar and axial channeling, whereas for positrons only one type of real channeling is known, the regime when the particle is localized in between two adjacent planes. Up to the present no solid researches of the possibility of axial channeling of positively charged particles has been conducted, because regardless of the crystal type the crystallographic axes are positively charged.

For radiation physics an examination of the possibility of axial channeling of positrons and, hence, the formation of metastable relativistic positron systems (PS) is a highly important problem. The channeling of positrons was dealt with in Ref. [14], where for the first time the observation of annihilation radiation from MeV energy positrons at transition through the crystal of gold was reported.
The macroscopic channeling in magnetic systems and in an intense electromagnetic wave have been studied in Refs. [15–22].

In the present work the process of annihilation of relativistic PS or of positron atom (PA), formed by the positron and an electron of medium, is analyzed and the lifetimes of positron in the channeling regime are obtained.

## 2. Formation of relativistic positron systems

As was shown in [23], in case of scattering at small angles to the axis <100> of chloride ion in CsCl crystals, low energy (5–20 MeV) relativistic positrons appear in the mode of axial channeling. Besides, the moving positrons are concentrated between two cylinders, that is highly important from the viewpoint of motion stability. It was shown that the effective 2D potential of channeling was independent of the temperature of medium in a wide temperature range and has a depth of order of 10 eV, that is sufficient for formation of several quantum states of the transverse motion.

Taking into account the symmetry of effective potential around the axis of negative ions for positrons, one can write down the following analytical formula for the effective potential:

$$U(\rho) = U_0 \left( e^{-2\alpha\bar{\rho}} - 2e^{-\alpha\bar{\rho}} \right), \quad \bar{\rho} = (\rho - \rho_0)/\rho_0, \quad \rho = \sqrt{x^2 + y^2}, \qquad (1)$$

where $U_0$ is the well depth, α determines the width of potential, ρ0 is the equilibrium distance. For ordinary crystals these parameters are within the intervals $U_0 = 5\text{-}10\,\text{eV}$, $\alpha = 0.5 - 0.8$ and $\rho_0 \sim 0.5d$, where *d* is the lattice constant.

The complete wave function of positron in (1) potential and eigenvalues of localized states are found and represented in Ref. [23].

## 3. processes leading to the decay of positron systems

After exclusion of main dechanneling factors, there still remain two different processes leading to the decay of PS: 1) annihilation of positron with an electron of medium into one gamma - photon; 2) annihilation of positron with an electron of medium into two gamma - photons. Our main task is the investigation of the lifetime of PS.

*3.1. Annihilation of Positron with Electron of Medium into One gamma – Photon*

It is evident that PS may be represented as a system with common zero spin (like the parapositronium) because the skeleton of negatively charged ionic axes has no spin and, respectively, the spin-spin interaction between them and the positron is absent, i.e. the positron and axis interact only electromagnetically. Another difference between the positronium and PS consists in the possibility of the one gamma - photon decay of the latter. In this case the laws of conservation of energy and momentum are met due to the presence of medium. This process is defined by the first-order matrix element [24]

$$Q = \langle f|S^{(1)}|i\rangle = \frac{2\pi i e}{\sqrt{2\omega}} \int \Psi^*(\mathbf{r}) \mathbf{e} e^{-i\mathbf{q}\mathbf{r}} \psi(\mathbf{r}) d^3 r \delta(\varepsilon_p + \varepsilon_e - \omega), \qquad (2)$$

where $e$ is the electron charge, $\Psi$ and $\varepsilon_p$ the wave function and total energy of positron, $\psi$ and $\varepsilon_e$ the wave function and energy of an electron of medium, $\mathbf{q}$ the momentum of gamma - photon, $\omega$ its frequency and $\mathbf{e}$ the unit vector of photon polarization.

The effective differential cross-section of annihilation may be written as follows:

$$d\sigma = \frac{e^2}{2(2\pi)^2 \omega} \sum_{\upsilon_e, \upsilon_p} |Q|^2 \delta(\varepsilon_p + \varepsilon_e - \omega) \omega^2 do_\gamma, \qquad (3)$$

where $do_\gamma = \sin\theta d\theta d\varphi \, do\gamma$ is an element of solid angle, within which the photon momentum is contained, $\upsilon_e$ and $\upsilon_p$ are the summation indices of the spins of electrons and positrons in the initial state.

Taking advantage of the procedure described in Ref. [23], we obtain for $Q$

$$Q = \frac{2(2\pi)^{5/2}}{\mu d} \delta(p_z - q_z) \hat{\psi}(0) p_z^{1/2} \int_0^\infty \Theta_0(p_\perp') p_\perp'^{3/2} dp_\perp', \qquad (4)$$

where $\hat{\psi}(0)$ is the Fourier transform of the wave function of electron and function $\Theta_0(p_\perp)$ is defined in Ref. [23]. Using expression (4) and the standard relation between the amplitude and probability [24] of one photon transition of PS decay, the probability is estimated as

$$P_\gamma \sim 10^6 \text{сек}^{-1}. \qquad (5)$$

## 3. Conclusion

Our investigations show that the problem connected with the shortness of dechanneling length is solvable if the channeling of positrons with 5–20 MeV energy is considered particularly in CsCl type ionic crystals along the chloride ions ( 100 axis). In this case a two-dimensional relativistic PS is formed that practically does not interact with the phonons of lattice subsystem. All the other types of action on PS, namely: collisions with electrons of medium, scattering on discretenesses of the lattice etc. are perturbations.